\documentclass[journal,letterpaper]{IEEEtran}

\usepackage[utf8]{inputenc} 
\usepackage[T1]{fontenc}
\usepackage{url}
\usepackage{ifthen}
\usepackage{cite}
\usepackage[cmex10]{amsmath} %

\usepackage[ruled]{algorithm}
\usepackage{colortbl}
\usepackage{booktabs}
\usepackage{multirow}
\usepackage{caption}
\usepackage{float}
\usepackage{multicol}
\usepackage[normalem]{ulem}
\usepackage{algpseudocode}
\usepackage{tikz-layers}
\algnewcommand{\Initialize}{%
  \State \textbf{Initialize:}
}
\algnewcommand{\Output}{%
  \State \textbf{Output:}
}
\usepackage{mathtools}
\usepackage{graphicx}
\usepackage{url}

\usepackage[
    shortcuts,       %
    acronym,         %
    section=section, %
]{glossaries}
\usepackage{tikz}
\usetikzlibrary{backgrounds,calc,shadings,shapes.arrows,shapes.symbols,shadows,fit,positioning,topaths,hobby,arrows.meta, decorations.pathmorphing}
\usepackage{float}
\usepackage{amssymb}
\usepackage{amsmath}
\usepackage{amsthm}
\usepackage{xspace}
\usepackage[capitalize]{cleveref}
\usepackage{subfig}
\usepackage[inline]{enumitem}
\usepackage{cite}
\usepackage{siunitx}
\usepackage{dsfont}
\usepackage{balance}
\makeglossaries

\newtheorem{theorem}{Theorem}
\newtheorem{corollary}{Corollary}
\newtheorem{lemma}{Lemma}

\newtheorem{proposition}{Proposition}

\newtheorem{remark}{Remark}

\usepackage{pgfplots}
\pgfplotsset{compat=newest}

\crefname{paragraph}{paragraph}{paragraphs}
\Crefname{paragraph}{Paragraph}{Paragraphs}

\IEEEoverridecommandlockouts

\newcommand{\camready}[1]{\textcolor{black}{#1}}

\newcommand{\alphabet}{\ensuremath{\mathcal{X}}}
\newcommand{\rvc}{\ensuremath{\mathsf{C}}}
\newcommand{\pdfc}{\ensuremath{p_\rvc}}
\newcommand{\nrec}{\ensuremath{N}}
\newcommand{\rec}{\ensuremath{i}}
\newcommand{\rectwo}{\ensuremath{i^\prime}}
\newcommand{\rep}{\ensuremath{k}}
\newcommand{\repidx}[1][\rec]{\ensuremath{p_{#1, \rep}}}
\newcommand{\repidxrow}{\ensuremath{\sampleidxcommon_\rep}}
\newcommand{\repidxrowtmp}{\ensuremath{\sampleidxcommon_{\rep^\prime}}}
\newcommand{\repidxcol}[1][\rec]{\ensuremath{\ell_{#1, \rep}}}
\newcommand{\nrep}{\ensuremath{K}}
\newcommand{\rvi}[1][\rec]{\ensuremath{\mathsf{Y}_{#1}}}
\newcommand{\rvwi}[1][\rec]{\ensuremath{\mathsf{W}_{#1}}}

\newcommand{\rvwc}[1][\rec]{\ensuremath{\mathsf{W}_{\rvc}}}
\newcommand{\pdfi}[1][\rec]{\ensuremath{p}_{\rvi[#1]}}
\newcommand{\pdfjoint}[1][\rec]{\ensuremath{p}_{\rvi[1], \cdots, \rvi[\nrec]}}
\newcommand{\pdfjointtwo}[1][\rec]{\ensuremath{p}_{\rvi[1], \rvi[2]}}
\newcommand{\pdfjointtwocondc}[1][\rec]{\ensuremath{p}_{\rvi[1], \rvi[2] \vert \rvc}}
\newcommand{\pdfauxc}[1][\rec]{\ensuremath{p}_{\rvwc[#1]}}
\newcommand{\pdfauxi}[1][\rec]{\ensuremath{p}_{\rvwi[#1]}}
\newcommand{\pdfauxic}[1][\samplecommon]{\ensuremath{p}_{\rvwi \vert \rvc}}
\newcommand{\pdfauxicvar}[1][\samplecommon]{\ensuremath{p}_{\rvwi \vert \rvc=#1}}
\newcommand{\fracc}[1][\rec]{\ensuremath{\rho}_{\rvc}}
\newcommand{\fraci}[1][\tmpsamplec]{\ensuremath{\rho}_{\rec}^{#1}}
\newcommand{\isweights}{\ensuremath{\rho_{\rec}}}
\newcommand{\target}{\ensuremath{\mathsf{Q}}}

\newcommand{\pdft}{\ensuremath{ p_{\target}}}
\newcommand{\pdftapprox}{\ensuremath{\tilde{p}_{\target}}}

\newcommand{\pdftc}{\ensuremath{p_{\target_\rvc}}}
\newcommand{\pdftcondc}{\ensuremath{p}_{\target \vert \rvc}}
\newcommand{\pdfcondc}[1]{\ensuremath{p}_{\rvc_{#1} \vert \rvc_{\the\numexpr #1-1\relax}}}
\newcommand{\pdftcondcvar}[1][\commonvar]{\ensuremath{p}_{\target \vert \rvc = #1}}

\newcommand{\function}{\ensuremath{f}}

\newcommand{\nsamplesindiv}[1][\rec]{\ensuremath{n_{#1,c}}}
\newcommand{\nsamplesindivvar}[1][\commonvar]{\ensuremath{n_{\rec, #1}}}
\newcommand{\nsamplesrec}[1][\rec]{\ensuremath{n_{#1}}}
\newcommand{\nsamplescommon}{\ensuremath{n_c}}
\newcommand{\nsamples}{\ensuremath{n}}
\newcommand{\sampleidxcommon}{\ensuremath{m}}

\newcommand{\sampleidx}{\ensuremath{j}}
\newcommand{\sampleindiv}[1][\sampleidx]{\ensuremath{Y_{\rec, #1}}}
\newcommand{\sampleindivstd}[1][\sampleidx]{\ensuremath{Y_{\rec, #1}}}
\newcommand{\sampleindivonestd}[1][\sampleidx]{\ensuremath{Y_{1, #1}}}
\newcommand{\sampleindivtwostd}[1][\sampleidx]{\ensuremath{Y_{2, #1}}}
\newcommand{\sampleindivone}[1][\sampleidx]{\ensuremath{Y_{1, #1}}}
\newcommand{\sampleindivtwo}[1][\sampleidx]{\ensuremath{Y_{2, #1}}}
\newcommand{\sampleindivcond}[2]{\ensuremath{Y_{\rec, #1}^{(#2)}}}
\newcommand{\sampleindivcondone}[2]{\ensuremath{Y_{1, #1}^{(#2)}}}
\newcommand{\sampleindivcondtwo}[2]{\ensuremath{Y_{2, #1}^{(#2)}}}
\newcommand{\samplecommon}[1][\sampleidx]{\ensuremath{C_{#1}}}
\newcommand{\is}[1][\function]{\ensuremath{I(#1)}}
\newcommand{\isest}{\ensuremath{I_\rec(\function)}}
\newcommand{\isesti}[1][\function]{\ensuremath{I_{\rec, \nrep}(\function)}}

\newcommand{\define}{\ensuremath{\triangleq}}
\newcommand{\tmpvarint}{\ensuremath{y}}
\newcommand{\kl}[2]{\ensuremath{D_{\text{KL}}(#1\Vert #2)}}
\newcommand{\tv}[2]{\ensuremath{d_{\text{TV}}(#1\Vert #2)}}
\newcommand{\cgk}[2]{\ensuremath{C_{\text{GK}}(#1; #2)}}

\newcommand{\mutinf}[2]{\ensuremath{\mathrm{I}(#1; #2)}}
\newcommand{\entropy}[1]{\ensuremath{\mathrm{H}(#1)}}
\newcommand{\gkfunctions}[1][\rec]{\ensuremath{g_{#1}}}
\newcommand{\pdficondc}[1][\rec]{\ensuremath{p}_{\rvi[#1]\vert \rvc}}
\newcommand{\pdfccondi}[1][\rec]{\ensuremath{p}_{\rvc \vert \rvi[#1]}}

\newcommand{\pdficondcvar}[1][\commonvar]{\ensuremath{p}_{\rvi[\rec]\vert \rvc=#1}}

\newcommand{\Ed}[2]{\mathbb{E}_{#1}\left[ #2 \right]}

\newcommand{\lipschitz}{\ensuremath{L}}
\newcommand{\consti}{\ensuremath{\tau_{\rec}}}
\newcommand{\flip}[1][\pdft]{\ensuremath{\Vert \function \Vert_{\lipschitz^2(#1)}}}

\newcommand{\gcondlip}[2]{\ensuremath{\Vert #1 \Vert_{\lipschitz^2(#2)}}}
\newcommand{\fournorm}[2]{\ensuremath{\Vert #1 \Vert_{\lipschitz^4(#2)}}}
\newcommand{\pnorm}[2]{\ensuremath{\Vert #1 \Vert_{\lipschitz^p(#2)}}}
\newcommand{\slack}{\ensuremath{t}}

\newcommand{\commonvar}{\ensuremath{c}}

\newcommand{\rvqc}{\ensuremath{\mathsf{Q_C}}}
\newcommand{\rvt}{\ensuremath{\mathsf{X}}}

\newcommand{\chisq}[2]{\ensuremath{\chi^2\!\left(#1,#2\right)}}

\newcommand{\efrac}{\ensuremath{\chisq{\pdfc}{\pdftc}+1}}

\newcommand{\maxblockeps}[1][\rec]{\ensuremath{\hat{\epsilon}_{#1}}}
\newcommand{\blockalphabet}{\ensuremath{\mathcal{C}}}

\newcommand\numberthis{\addtocounter{equation}{1}\tag{\theequation}}

\begin{document}
\title{Multi-Terminal Remote Generation and Estimation Over a Broadcast Channel With Correlated Priors} 

\author{%
    \IEEEauthorblockN{Maximilian Egger, %
                    Rawad Bitar, %
                    Antonia Wachter-Zeh, %
                    Nir Weinberger and %
                    Deniz Gündüz \vspace{-.5cm}%
                    } %
    \thanks{M.E., R.B. and A.W-Z. are with the Technical University of Munich. Emails: \{maximilian.egger, rawad.bitar, antonia.wachter-zeh\}@tum.de. D.G. is with Imperial College London. Email: d.gunduz@imperial.ac.uk. N.W. is at Technion --- Israel Institute of Technology. Email: nirwein@technion.ac.il.}
    \thanks{This project has received funding from the German Research Foundation (DFG) under Grant Agreement Nos. BI 2492/1-1 and WA 3907/7-1, and from UKRI for project AI-R (ERC-Consolidator Grant, EP/X030806/1). The work of N.W. was partly supported by the Israel Science Foundation (ISF), grant no. 1782/22.}
}

\maketitle

\begin{abstract} 
We study the multi-terminal remote estimation problem under a rate constraint, in which the goal of the encoder is to help each decoder estimate a function over a certain distribution -- while the distribution is known only to the encoder, the function to be estimated is known only to the decoders, and can also be different for each decoder. The decoders can observe correlated samples from prior distributions, instantiated through shared randomness with the encoder. To achieve this, we employ remote generation, where the encoder helps decoders generate samples from the underlying distribution by using the samples from the prior through importance sampling. While methods such as minimal random coding can be used to efficiently transmit samples to each decoder individually using their importance scores, it is unknown if the correlation among the samples from the priors can reduce the communication cost using the availability of a broadcast link. We propose a hierarchical importance sampling strategy that facilitates, in the case of non-zero Gács-Körner common information among the priors of the decoders, a common sampling step leveraging the availability of a broadcast channel. This is followed by a refinement step for the individual decoders. We present upper bounds on the bias and the estimation error for unicast transmission, which is of independent interest. We then introduce a method that splits into two phases, dedicated to broadcast and unicast transmission, respectively, and show the reduction in communication cost.
\end{abstract}

\section{Introduction}

In machine learning (ML), learning generative distributions to model data in some domain has become a prominent discipline. In our problem, an entity -- hereby referred to as the encoder -- aims to allow a group of decoders to estimate statistical properties of its distribution, such as population inferences derived from ML models \camready{(e.g., in generative adversarial networks \cite{goodfellow2014generative} or variational auto-encoders \cite{kingma2013auto})}. The distribution is known only to the encoder, while each decoder may want to estimate a different function of that distribution. A direct solution for the encoder is to sample from its distribution and transmit deterministic samples to decoders. However, this method can lead to substantial communication overheads, introduce considerable latency in remote estimation, and potentially compromise the privacy of the encoder’s distribution. \camready{Similarly, compressing and sending the distribution directly is suboptimal unless the distribution belongs to a simple class.} To address these issues, prior distributions available at the decoders (also known to the encoder) can be used to remotely generate samples, which can then be used to estimate the desired statistics through \textit{importance sampling} \cite{chatterjee2018sample}. In this multi-terminal setting, the use of a broadcast channel from the encoder to the decoders can exploit potential correlations of samples observed from the decoders' prior distributions, in order to reduce the communication costs. In this paper, we explore the gains of this approach.

A point-to-point remote version of sample generation and estimation was considered in \cite{havasi2018minimal}. A scheme called minimal random coding  (MRC) was proposed, which is a lossy compression method. The scheme is based on selecting a sample from a proposal distribution using the relative importance of scores with respect to the target distribution. In MRC, the encoder and the decoder draw the same samples from the prior distribution using shared randomness. The encoder then draws an index of one of these samples from a distribution that reflects the relative importance of the sample with respect to the target distribution. This problem is mathematically equivalent to channel simulation (also known as reverse channel coding) \cite{theis2022algorithms,li2024channel,flamich2024some} and relative entropy coding (REC) \cite{flamich2020compressing}, where the general narrative is to make a decoder sample from a distribution that is close or equal to the target distribution. Closely related is also the function representation lemma \cite{li2018strong}. REC was initially proposed for image compression, with various follow-up works aiming to improve its complexity, e.g., \cite{flamich2022fast,flamich2024faster,he2024accelerating}. %
In parallel, MRC was improved by ordered random coding (ORC) \cite{theis2022algorithms}, thereby connecting importance sampling with Poisson functional representation. Building on ORC, importance sampling-based lossy compression was further developed in \cite{phan2024importance}.

The aforementioned works have recently attracted interest in ML, as they allow for alleviating various negative impacts of quantization in these systems \cite{chen2024information}. For example, in \cite{triastcyn2021dp}, a combination of REC and differential privacy was used for communication-efficient federated learning. In \camready{\cite{isik2024adaptive,egger2025bicompfl}, general frameworks for stochastic federated learning were proposed, leveraging MRC} for communication-efficient transmission of samples from the clients' posteriors, i.e., their model updates. In this setting, multiple clients transmit importance samples to the federator\camready{, and proposal distributions capture temporal correlation in the learning process.}

With the exception of \cite{isik2024adaptive}, most works study the case of a single encoder and a single decoder, whereas the multi-terminal setting, and specifically, multiple decoders, has received much less attention. However, this setting is particularly interesting when the decoders can observe \textit{correlated} samples from their joint prior distribution. Naively, using shared randomness between encoder and decoder, the encoder can perform MRC to convey to each decoder samples from its respective prior distribution, reflecting the relative importance with respect to the target distribution. The communication cost of this strategy scales linearly in the number of decoders. However, such a large cost can be reduced in the presence of a broadcast channel. For instance, in the extreme case where the joint prior is such that the samples are fully correlated (equal), it suffices for the encoder to transmit just a single sample index to all the decoders, via a broadcast channel. Our goal in this paper is to characterize the potential gains in the case of joint prior distributions with arbitrary correlation. We focus on a specific instance of the problem, where the samples obtained by the decoders are used to evaluate a function of interest. %

Our contributions are as follows. As a preliminary step, we refine a known result on the sampling process in MRC, and provide in \cref{sec:prelims} a high-probability error bound for the standard procedure in the single-decoder case. Then, in \cref{sec:nested_sampling}, we propose a hierarchical sampling strategy that operates in two stages: It first samples a block from the support of the prior distribution, and second, conditionally samples the final index within the chosen block. We analyze the bias and the error incurred in the function evaluation using this sampling strategy, the transmission cost, and the distance of the final sample distribution to the target distribution. The results are compared to non-hierarchical (standard) sampling procedures. Then, using the Gács-Körner common information shared by two decoders, we apply hierarchical sampling in the multi-decoder setting, and establish the reduction in communication cost with a broadcast channel, followed by decoder-specific refinements through unicast transmission. %

\section{Preliminaries and System Model} \label{sec:prelims}

\textbf{Notation.} We denote the Kullback-Leibler divergence between two distributions $p_\mathsf{U}$ and $p_{\mathsf{V}}$ by $\kl{p_\mathsf{U}}{p_{\mathsf{V}}}$, the total variation distance by $\tv{p_\mathsf{U}}{p_{\mathsf{V}}}$, and the $\chi^2$-divergence by $\chisq{p_\mathsf{U}}{p_{\mathsf{V}}}$. For a natural number $u$, we let $[u] \define \{1, \cdots u\}$. We denote the entropy of $\mathsf{U}$ by $\entropy{\mathsf{U}}$, and the mutual information between $\mathsf{U}$ and $\mathsf{V}$ by $\mutinf{\mathsf{U}}{\mathsf{V}}$. The Gács-Körner common information $\mathsf{C}$ between $\mathsf{U}$ and $\mathsf{V}$ is given by $\cgk{\mathsf{U}}{\mathsf{V}} = \max_{\mathsf{C}} \entropy{\mathsf{C}}$ s.t. $\mathsf{C} = \gkfunctions[1](\mathsf{U}) = \gkfunctions[2](\mathsf{V})$ for two deterministic functions $\gkfunctions[1]$ and $\gkfunctions[2]$. We denote the $p$-norm of a function $h(\cdot)$ over distribution $p_\mathsf{u}$ as $\pnorm{h}{p_\mathsf{u}} \define \Ed{\mathsf{U} \sim p_\mathsf{u}}{\vert h(\mathsf{U}) \vert^p}^{1/p}$.

We consider the setting of a single encoder and multiple decoders $\rec \in [\nrec]$. The encoder shares randomness with the decoders $\rec \in [\nrec]$, each of which can observe samples from a prior distribution $\pdfi$ induced by a joint distribution $\pdfjoint$. \camready{Following the common assumption,} the joint distribution $\pdfjoint$ is known to all parties, yet decoder $\rec$ can only observe samples from the marginal $\pdfi$, and not from $\pdfi[\rec^\prime]$ for $\rectwo \in [\nrec] \setminus \{\rec\}$. According to the joint distribution, the samples drawn from the priors $\pdfi$ at the different decoders might be arbitrarily dependent.

\begin{figure}[!t]
    \centering
    \resizebox{\linewidth}{!}{
    \begin{tikzpicture}[node distance=1.5cm and 3.5cm, >=Stealth, thick]

\node[draw, rectangle, minimum width=2cm, minimum height=1cm, align=center, inner sep=2pt] (encoder) {Encoder\\ with $\pdft$};

\node[draw, rectangle, minimum width=2cm, minimum height=1cm, align=center, right=2cm of encoder, yshift=1.3cm, inner sep=2pt] (receiver1) {Decoder 1};
\node[draw, rectangle, minimum width=2cm, minimum height=1cm, align=center, right=2cm of encoder, yshift=-1.3cm, inner sep=2pt] (receiver2) {Decoder 2};

\draw[->, thick] (encoder.east) -- ++(.5, 0) coordinate (midpoint) |- node[midway, above right, text=black] {$\mathbf{m}$} (receiver1.west);
\draw[->, thick] (midpoint) |- node[midway, below right, text=black] {$\mathbf{m}$} (receiver2.west);

\draw[->, dashed, thick] (encoder.east) to[out=45, in=180] node[midway, below, sloped, text=black] {$\mathbf{m}_1$} (receiver1.west);
\draw[->, dashed, thick] (encoder.east) to[out=-45, in=180] node[midway, above, sloped, text=black] {$\mathbf{m}_2$} (receiver2.west);

\node[below=0.4cm of encoder] (sample_encoder) {\small $Y_{1,j}, Y_{2,j} \sim p_{Y_1, Y_2}$};
\draw[-, dashed] (encoder.south) -- (sample_encoder.north);

\node[below=0.4cm of receiver1] (sample_receiver1) {\small $Y_{1,j} \sim p_{Y_1}$};
\draw[-, dashed] (receiver1.south) -- (sample_receiver1.north);

\node[below=0.4cm of receiver2] (sample_receiver2) {\small $Y_{2,j} \sim p_{Y_2}$};
\draw[-, dashed] (receiver2.south) -- (sample_receiver2.north);

\node[draw, rectangle, minimum width=3cm, minimum height=1cm, align=center, right=1cm of receiver1, xshift=1.5cm, inner sep=2pt] (function1) {$\frac{1}{\nrep} \sum_{\rep=1}^\nrep \function(X_1^{(\rep)})$};
\node[draw, rectangle, minimum width=3cm, minimum height=1cm, align=center, right=1cm of receiver2, xshift=1.5cm, inner sep=2pt] (function2) {$\frac{1}{\nrep} \sum_{\rep=1}^\nrep \function(X_2^{(\rep)})$};

\draw[->, thick] (receiver1.east) -- (function1.west) node[midway, above, text=black] {$X_1^{(1)}\!\!\!, \cdots\!, \! X_1^{(\nrep)}$};
\draw[->, thick] (receiver2.east) -- (function2.west) node[midway, above, text=black] {$X_2^{(1)}\!\!\!, \cdots\!, \! X_2^{(\nrep)}$};

\end{tikzpicture}}
    \vspace{-.6cm}
    \caption{System model for two decoders. \vspace{-.4cm}} %
    \vspace{-.2cm}
    \label{fig:setting}
\end{figure}
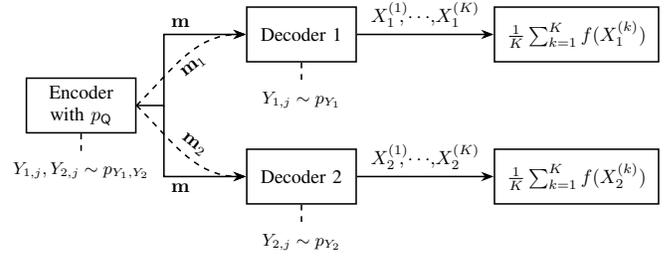

The encoder has access to a \camready{random variable $\target$}, defining a target probability measure $\pdft$ over $\alphabet$, where $\pdft$ is unknown to the decoders. Let $\function_{\rec}: \alphabet \rightarrow \mathbb{R}, \forall \rec \in [\nrec],$ be arbitrary measurable functions over an alphabet $\alphabet$, where $\function_\rec$ is only available to decoder $\rec$. For each $\rec$, the encoder wants to estimate the evaluation of the functions $\function_\rec$ over the target measure $\pdft$, i.e.,
$ \is[\function_\rec] \define \int_\alphabet \function_\rec(\tmpvarint) d \pdft(\tmpvarint).$ 
The encoder, having observations $Y_{1,j}, \cdots, Y_{\nrec, j} \sim \pdfjoint$, sends a message $\mathbf{m}$ to all decoders over a rate-constraint broadcast channel, and transmits a message $\mathbf{m}_\rec$ to decoder $\rec$ over a rate-constraint unicast channel. Potentially using the observations $\sampleindiv$ from the prior $\pdfi$ shared with the encoder \camready{and the message $\mathbf{m}_\rec$}, each decoder $\rec$ reconstructs samples $X_\rec^{(\rep)}, \forall \rep \in [\nrep],$ for some $\nrep>0$, to be used as the input to the function $\function_\rec$, thereby obtaining an estimate $\isesti[\function_\rec] \define \frac{1}{\nrep} \sum_{\rep = 1}^\nrep \function_\rec(X_\rec^{(\rep)})$. 
Although our results apply to the case where the function is different for each of the decoders, we assume for clarity the same function $\function$ at all decoders. The system model is depicted in \cref{fig:setting}.

The direct method for estimating $\is$ by the decoder is for the encoder to sample from $\pdft$, and then transmit deterministic quantized samples to each decoder, which allows them to estimate $\is$. However, the communication cost can be significantly reduced by leveraging the prior information $\pdfjoint$, known to both the encoder and the decoders, especially when the prior distributions $\pdfi$ are close to the target distribution $\pdft$. We use a channel simulation approach, where, using the priors $\pdfi$, we make the decoders sample from the target distribution $\pdft$, or a distribution $\pdftapprox \approx \pdft$. Those samples are then used to estimate the function evaluation.

A naive solution is to treat each of the decoders $\rec \in [\nrec]$ individually. Following this solution, let us consider for now one of the decoders $\rec$, and utilize the MRC method \cite{havasi2018minimal}, which does not require or utilize the knowledge of $\pdft$ at the decoder. %
The encoder draws $\nsamples$ samples $\sampleindiv[1], \cdots, \sampleindiv[\nsamples] \sim \pdfi$ and constructs a categorical auxiliary distribution $\pdfauxi$ as \vspace{-.1cm}
\begin{align*}
\pdfauxi(\sampleidx) = \frac{\pdft(\sampleindiv)/\pdfi(\sampleindiv)}{\sum_{\sampleidx=1}^\nsamples \pdft(\sampleindiv)/\pdfi(\sampleindiv)}. \vspace{-.2cm}
\end{align*}
The encoder then draws an index $\repidx \sim \pdfauxi$ and transmits it to the decoder $\rec$ using $\log_2(\nsamples)$ bits. The decoder $\rec$ reconstructs the sample $\sampleindiv[\repidx]$ by drawing the same samples $\sampleindiv[1], \cdots \sampleindiv[\nsamples] \sim \pdfi$ using the common prior and shared randomness\footnote{Improved REC strategies were proposed in \cite{flamich2020compressing,flamich2022fast,flamich2024faster,theis2022algorithms}. However, for simplicity, we focus here on the MRC strategy introduced in \cite{havasi2018minimal}.}.

By repeating this process for multiple samples $\rep \in [\nrep]$ and separately for each of the decoders, each decoder $\rec \in [\nrec]$ obtains samples used to estimate the function's value as $
    \isesti \define \frac{1}{\nrep} \sum_{\rep = 1}^\nrep \function(\sampleindiv[\repidx]). %
$
Letting $\consti \define \sum_{\sampleidx=1}^\nsamples \pdft(\sampleindiv)/\pdfi(\sampleindiv)$, we obtain  the following expression for the expected value of the estimate $\isesti$: \vspace{-.07cm}
\begin{align}
    \!\!\!\Ed{\repidx \sim \pdfauxi}{\function(\sampleindiv[\repidx])} = \frac{1}{\consti} \sum_{\repidx \in [\nsamples]} \frac{\pdft(\sampleindiv[\repidx])}{\pdfi(\sampleindiv[\repidx])} \function(\sampleindiv[\repidx]). \label{eq:exp_mrc}
\end{align} 
The analysis of importance sampling in  \cite{chatterjee2018sample} states that an accurate estimate of $\is$ can be obtained by observing $\nsamples = \mathcal{O}(\exp(\kl{\pdft}{\pdfi}))$ samples $\sampleindiv[1], \cdots, \sampleindiv[\nsamples]$ from $\pdfi$ and normalizing by the importance weights, i.e., 
$
    \isest = \frac{1}{\nsamples} \sum_{\sampleidx = 1}^\nsamples \function(\sampleindiv[\sampleidx]) \isweights,
$
where $\isweights \define \frac{\pdft(\sampleindiv[\sampleidx])}{\pdfi(\sampleindiv[\sampleidx])}$. Up to a normalization constant, this estimate corresponds to the expectation of the MRC approach given in \eqref{eq:exp_mrc}. 
Hence, the error bound of the importance-sampling estimate in \cite{chatterjee2018sample} directly leads to a bound on the bias of the MRC approach incurred by the sampling: %
\begin{lemma}[\!\!\!\!\!\protect{\cite[Theorem 2.1]{chatterjee2018sample}}] \label{lemma:bias}
For some $\slack \geq 0$, let $\nsamples = \exp(\kl{\pdft}{\pdfi} + \slack)$ and  $\epsilon \define e^{-\slack/4} + 2 \sqrt{\Pr_{X \sim \pdft}(\log(\fraci[](X) > \kl{\pdft}{\pdfi} + \slack/2))})^{1/2}$, then with probability at least $1-2\epsilon$, \vspace{-.4cm}
\begin{align*}
\vert \Ed{\repidx \sim \pdfauxi}{\function(\sampleindiv[\repidx])} - \Ed{X \sim \pdft}{\function(X)} \vert \leq 2\frac{\flip[\pdft] \epsilon}{1-\epsilon},\\[-.7cm]
\end{align*}
where the probability is over the randomness of $\sampleindiv \sim \pdfi$.
\end{lemma}

Using this result on the bias of the sampling process, along with concentration of the samples of the estimator, we derive the following guarantee on our estimator $\isesti$:
\begin{proposition} \label{prop:dev} Let $\nsamples$ and $\epsilon$ be defined as in \cref{lemma:bias}. For some $\epsilon^\star > 0$, with probability at least $1-\epsilon^\star-4\epsilon$, we have \vspace{-.35cm}
\begin{align*}
    &\vert \isesti \!\! - \!\! \is \vert \!\leq \!\! \frac{2\epsilon\flip[\pdft]}{1-\epsilon} \! + \! \frac{\sqrt{\!\frac{2\epsilon}{1-\epsilon}} \fournorm{\function}{\pdft} \!\! + \!\! \sqrt{\!\gcondlip{\function}{\pdft}}}{\nrep \epsilon^\star}\!.\\[-.7cm]
\end{align*}
\end{proposition}

For each decoder, the number of samples $\nsamples$ required for low variance of the estimator is given by $\mathcal{O}(\exp(\kl{\pdft}{\pdfi}))$, and this cost scales linearly with the number of decoders. 
Whenever only a unicast link is available, it appears that this cannot be improved. However, when a broadcast link is available, and the joint prior distribution is of correlated samples, the cost compared to the unicast case can be reduced. As discussed above, in the extreme case of equal and fully dependent $\pdfi$, broadcasting the same index to all decoders is trivially effective, but the solution is much less obvious for joint distributions with partial correlation. %
We propose to gauge the correlation using the  Gács-Körner common information and develop a method that reduces the communication cost, utilizing the availability of a broadcast channel.

\section{Hierarchical Sampling} \label{sec:nested_sampling}

\renewcommand{\sampleindiv}[1][\sampleidx]{\ensuremath{\sampleindivcond{\protect{#1}}{\protect{\samplecommon[\repidxrow]}}}}
\renewcommand{\sampleindivone}[1][\sampleidx]{\ensuremath{\sampleindivcondone{\protect{#1}}{\protect{\samplecommon[\repidxrow]}}}}
\renewcommand{\sampleindivtwo}[1][\sampleidx]{\ensuremath{\sampleindivcondtwo{\protect{#1}}{\protect{\samplecommon[\repidxrow]}}}}

For methodological reasons, we introduce the hierarchical sampling approach for a point-to-point setting, where the goal is to make one decoder $\rec \in [\nrec]$ estimate $\is$ by leveraging its prior $\pdfi$. 
In the following section, we will show how this technique can reduce the communication cost for multiple decoders that are required to estimate the function $\is$, and their samples follow a joint prior distribution with non-zero Gács-Körner common information.

\vspace{-.5cm}

\subsection{Hierarchical Sampling Methodology}
Let $\gkfunctions(\cdot)$ describe a mapping from the alphabet $\alphabet$ to a smaller alphabet $\blockalphabet$ that partitions alphabet $\alphabet$ into $\vert \blockalphabet \vert$ disjoint blocks. Hence, $\gkfunctions(\rvi)$ induces a random variable $\rvc = \gkfunctions(\rvi)$ with probability law $\pdfc$ determined by disjoint blocks of $\pdfi$. Let $\pdficondc$ describe the conditional distribution over a subset of $\alphabet$ determined by the choice of the block $\commonvar \in \blockalphabet$, such that $\pdficondc \pdfc = \pdfccondi \pdfi = \pdfi$. With $\gkfunctions(\cdot)$, we can define a random variable $\rvqc = \gkfunctions(\rvt), \rvt \sim \pdft$, that describes the target distribution $\pdft$ over the blocks $\blockalphabet$. The corresponding probability measure is denoted $\pdftc$. Similarly, we let the conditional distribution based on the block choice be $\pdftcondc$.

The hierarchical sampling scheme we propose is to perform MRC using $\pdfc$ to obtain a sample $\commonvar$ from $\rvc$ with high importance with respect to $\rvqc$, and then conduct conditional MRC using $\pdftcondcvar[\commonvar]$.

\subsubsection{Block Sampling} \label{paragraph:block_sampling}
The encoder draws $\nsamplescommon$ samples from $\pdfi$ and transforms them using $\gkfunctions(\cdot)$, thereby obtaining samples $\samplecommon[1], \cdots \samplecommon[\nsamplescommon] \sim \pdfc$ from $\pdfc$. The importance of the samples from $\pdfc$ w.r.t. $\pdftc$ is determined by the quantity $\fracc \define \frac{d \pdftc}{d \pdfc}$. 
The encoder constructs a categorical distribution $\pdfauxc$ reflecting the normalized importance scores $\fracc$ of the samples $\samplecommon, \sampleidx \in [\nsamplescommon]$, i.e.,\vspace{-.2cm}
\begin{align}
\pdfauxc(\sampleidx) = \frac{\pdftc(\samplecommon)/\pdfc(\samplecommon)}{\sum_{\sampleidx=1}^{\nsamplescommon} \pdftc(\samplecommon)/\pdfc(\samplecommon)}. \label{eq:pdfauxc} \\[-.7cm] \nonumber
\end{align}
The encoder then draws $\nrep$ indices $\repidxrow, \rep \in [\nrep]$ from the weight distribution $\pdfauxc$ and transmits those indices to decoder $\rec$ with $\nrep \log_2(\nsamplescommon)$ bits. 
\subsubsection{Conditional Sampling} \label{paragraph:conditional_sampling} The following is repeated for all $\rep \in [\nrep]$. Having selected an index $\repidxrow$, and hence a sample $\samplecommon[\repidxrow]$, the encoder draws samples from $\pdfi$ until obtaining $\nsamplesindivvar[\protect{\samplecommon[\repidxrow]}]$ samples $\sampleindiv[1], \cdots, \sampleindiv[\protect{\nsamplesindivvar[\protect{\samplecommon[\repidxrow]}]}] \sim \pdficondcvar[\protect{\samplecommon[\repidxrow]}]$ for which it holds that $\gkfunctions(\sampleindiv) = \samplecommon[\repidxrow]$ for all $\sampleidx \in [\nsamplesindivvar[\protect{\samplecommon[\repidxrow]}]]$. For each $\commonvar \in \blockalphabet$, we refer to the conditional importance as $\fraci \define \frac{d \pdftcondc(\cdot \vert \commonvar)}{d \pdficondc(\cdot \vert \commonvar)}$. The encoder constructs a second categorical distribution reflecting the conditional importance scores\footnote{The distribution $\pdfauxicvar[\protect{\samplecommon[\repidxrow]}]$ is reused for equal block samples, i.e., when $\samplecommon[\repidxrow]=\samplecommon[\repidxrowtmp]$ for $\rep \neq \rep^\prime \in [\nrep]$.}, i.e., setting $\commonvar = \samplecommon[\repidxrow]$,
\vspace{-.3cm}
\begin{align}
\pdfauxicvar[\commonvar](\sampleidx) = \frac{\pdftcondc(\sampleindivcond{\sampleidx}{\commonvar} \vert \commonvar)/\pdficondc(\sampleindivcond{\sampleidx}{\commonvar} \vert \commonvar)}{\sum_{\sampleidx=1}^{\nsamplesindivvar[\protect{\commonvar}]} \pdftcondc(\sampleindivcond{\sampleidx}{\commonvar} \vert \commonvar)/\pdficondc(\sampleindivcond{\sampleidx}{\commonvar} \vert \commonvar)}. \label{eq:pdfauxic} \\[-.7cm] \nonumber
\end{align}
The encoder draws an index $\repidxcol \sim \pdfauxicvar[\protect{\samplecommon[\repidxrow]}]$, and transmits it with $\log_2(\nsamplesindivvar[\protect{\samplecommon[\repidxrow]}])$ bits. For each $\rep \in [\nrep]$, the decoder thus reconstructs the sample $\sampleindiv[\repidxcol]$. 
The resulting function estimate is given by
$%
    \isesti = \frac{1}{\nrep} \sum_{\rep = 1}^\nrep \function(\sampleindiv[\repidxcol]). 
$ %
Uniquely determining a sample requires $\log_2(\nsamplescommon) + \log_2(\nsamplesindiv)$ bits to transmit the respective indices, where $\commonvar$ is the block sample selected in the first round. The block index $\repidxrow$ corresponds to a rough sample, and $\repidxcol$ describes the ``refinement'' of the sample for the prior $\pdfi$ w.r.t. the common part $\pdfc$. 
We thus refer to this strategy as ``hierarchical sampling''. 
It is not obvious whether the proposed strategy is inferior to the standard MRC procedure described in \cref{sec:prelims}. Thus, we next analyze the performance.

\begin{remark}
We note that the communication cost to transmit the indices $\repidxrow$ and $\repidxcol$ can be further reduced by known improved strategies leveraging the non-uniformity of the indices, cf. ORC \cite{theis2022algorithms}. We omit such improvements, as they distract from the main theme. The functionality and improvements brought by the scheme presented in \cref{sec:multi_terminal} remain unaffected.
\end{remark}

\subsection{On the Bias and Estimation Error}
First, we derive an upper bound on the bias of the hierarchical sampling procedure, followed by a bound on the average communication complexity. 
For clarity in the statement of the theorem, we define the following quantities.
For $\slack_c \geq 0$, let \vspace{-.01cm}
$$\epsilon^2(\slack_c) \! \define e^{-\frac{\slack_c}{4}} \! + \! 2 \sqrt{\!\!\!\!\!\!\!\Pr_{\,\,\,\,\,\,\,\,\rvqc \sim\pdftc}\!\!\!\!(\log(\fracc(\rvqc) \!\! > \!\! \kl{\pdftc}{\pdfc} \! + \! \slack_c/2))}.\vspace{-.3cm}$$ 
Further, for $\slack \geq 0$ and $d_c \define \kl{\pdftcondcvar[\commonvar]}{\pdficondcvar[\commonvar]}$, we have \vspace{-.15cm}
$$\epsilon^2_{\rec}(\commonvar, \slack) \define e^{-\frac{\slack}{4}} + 2 \sqrt{\Pr_{ X\sim\pdftcondcvar[\protect{\commonvar}]}(\log(\fraci[\commonvar](\rvt) > d_c \! + \! \slack/2))}.\vspace{-.2cm}$$
We let $\maxblockeps(\slack) \define \max_{\commonvar \in \blockalphabet} \epsilon_\rec(\commonvar, \slack)$.

\begin{theorem} \label{thm:bias_refinement}
For some fixed $\slack_c, \slack \geq 0$, let $\maxblockeps \define \maxblockeps(\slack)$, and $\epsilon \define \epsilon(\slack_c)$. When $\nsamplescommon \geq \exp(\kl{\pdftc}{\pdfc} + \slack_c)$ and $\forall \commonvar \in \blockalphabet: \nsamplesindiv \geq \exp(\kl{\pdftcondcvar[\commonvar]}{\pdficondcvar[\commonvar]} + \slack)$ we have with probability at least $1-2(\vert \blockalphabet \vert \maxblockeps+\epsilon)$ that
\begin{align*}
    &\big\vert \Ed{\repidxrow \sim \pdfauxc, \repidxcol \sim \pdfauxicvar[\protect{\samplecommon[\repidxrow]}]}{\function(\sampleindiv[\repidxcol])} - \Ed{X \sim \pdft}{\function(X)} \big\vert \\
    &\leq \flip \left[\frac{2\sqrt{2} \epsilon}{1-\epsilon} \left(\frac{2\maxblockeps}{1-\maxblockeps} + 1 \right) + \frac{2\maxblockeps}{1-\maxblockeps} \right] \numberthis \label{eq:bias_refinement} \\
    &\overset{(\star)}{\leq} 2\sqrt{2} \flip \left[\frac{\epsilon}{1-\epsilon} + \frac{\maxblockeps}{1-\maxblockeps} \right],
\end{align*}
where $(\star)$ holds for $\epsilon \leq \frac{1}{9}$.
\end{theorem}
Using \cref{thm:bias_refinement}, we can obtain an upper bound on the total variation distance between the distribution from which the sample is drawn, and the desired distribution $\pdft$.

\begin{corollary} \label{cor:total_variation}
Let $\pdftapprox$ be the distribution of the selected sample $\sampleindivcond{\repidxcol}{\samplecommon[\protect{\repidxrow}]}$. Under the assumptions stated in \cref{thm:bias_refinement} and for $\epsilon \leq \frac{1}{9}$, the total variation is bounded as
\begin{align*}
\tv{\pdftapprox}{\pdft} \leq 2(\vert \blockalphabet \vert+1) \maxblockeps + 4\epsilon.
\end{align*}
\end{corollary}

Following similar lines as in the proof of \cref{prop:dev}, we further obtain the following guarantee on the estimate $\isesti$.
\begin{proposition} \label{prop:dev_refinement}
Let $\nsamplescommon$, $\nsamplesindiv, \forall \commonvar \in \blockalphabet$, $\epsilon$ and $\maxblockeps$ be defined as in \cref{thm:bias_refinement}. For  $\epsilon^\star>0$ and $\hat{\epsilon} \define \left[\frac{2\sqrt{2} \epsilon}{1-\epsilon} \left(\frac{2\maxblockeps}{1-\maxblockeps} + 1 \right) + \frac{2\maxblockeps}{1-\maxblockeps} \right]$, we have with probability at least $1-\epsilon^\star-4(\vert \blockalphabet \vert \maxblockeps+\epsilon)$ that
\begin{align*}
    &\vert \isesti \! - \! \is \vert \!\leq \! \flip \hat{\epsilon} \! +\! \frac{\gcondlip{\function^4}{\pdft} \sqrt{\hat{\epsilon}} \! + \! \sqrt{\gcondlip{\function}{\pdft}}}{\nrep \epsilon^\star} \vspace{.7cm}
\end{align*}
\end{proposition}
We compare our bound in \eqref{eq:bias_refinement} to the established result in \cref{lemma:bias} in two edge cases: i) When each block contains only one symbol, the bound degenerates to $\flip \frac{2 \sqrt{2} \epsilon}{1-\epsilon}$. The additional factor of $\sqrt{2}$ is an artifact of the bounding technique. In fact, a slightly modified proof of \cref{thm:bias_refinement} can recover $\flip \frac{2 \epsilon}{1-\epsilon}$ in this special case; ii) When one block contains all symbols, the bound yields $\flip \frac{2 \maxblockeps}{1-\maxblockeps} = \flip \frac{2 \epsilon_\rec(1,\slack)}{1-\epsilon_\rec(1,\slack)}$, where in this case, $\epsilon_\rec(1,\slack)$ is defined over $\kl{\pdft}{\pdfi}$, i.e., $\epsilon_\rec(1,\slack) = e^{-\slack/4} + 2 \sqrt{\Pr_{\rvt \sim\pdft}(\log(\isweights(\rvt) > \kl{\pdft}{\pdfi} + \slack/2))})^{1/2}$, and $\isweights$ is defined as in \cref{sec:prelims}. Hence, we recover \cref{lemma:bias}. Interpolating between the edge cases i) and ii), i.e., when $\gkfunctions(\cdot)$ is non-trivial, the bound incurs a slight additional cost.

Despite the possible additional cost, the hierarchical sampling strategy \camready{is} beneficial over MRC even in the point-to-point setting, depending on the function properties. Indeed, consider the following illustrative example. 
In certain cases, sampling from $\pdftc$ with large precision is more important than sampling from $\pdftcondc$. Therefore, suppose that the function $\function$ of interest is smooth on the support of $\pdftcondc$ for all $\commonvar \in \blockalphabet$. In the most extreme case, the function has zero variance conditioned on a block. Then, an intermediate result of the proof of \cref{thm:bias_refinement} shows the following improved result.
\begin{corollary}
Let $\function$ be constant on the support of $\pdftcondc(\cdot \vert \commonvar)$ for all $\commonvar \in \blockalphabet$. 
For some fixed $\slack_c \geq 0$, let $\epsilon \define \epsilon(\slack_c)$. When $\nsamplescommon \geq \log(\kl{\pdftc}{\pdfc} + \slack_c)$ we have w.p. at least $1-2\epsilon$ that
\begin{align*}
&\big\vert \Ed{\repidxrow, \repidxcol}{\function(\sampleindiv[\repidxcol])} - \Ed{X \sim \pdft}{\function(X)} \big\vert \leq \flip \frac{2 \epsilon}{1-\epsilon}.
\end{align*}
\end{corollary}
This result significantly improves over \cref{lemma:bias} by requiring a lower number of samples $\nsamplescommon$ and providing a tighter probabilistic guarantee determined by $\epsilon$. 
Generalizing this concept to the case where $\function$ is smooth but not constant on the support of the conditionals $\pdftcondc$, i.e., where the index $\repidxcol$ carries only little information compared to the index $\repidxrow$, illustrates the potential of hierarchical sampling for certain point-to-point problems, e.g., successive refinement.

\vspace{-.3cm}

\subsection{Complexity Analysis}
We continue to analyze the cost of the proposed strategy. From \cref{thm:bias_refinement}, it is immediate that the sample complexity is upper bounded by $\mathcal{O}(\exp(\kl{\pdftc}{\pdfc})) + \max_{\commonvar \in \blockalphabet} \mathcal{O}(\exp(\kl{\pdftcondcvar[\commonvar]}{\pdficondcvar[\commonvar]}))$. However, based on $\pdfc$ and $\pdftc$ and their divergence $\chisq{\pdfc}{\pdftc}$, we can obtain a tighter guarantee on the average complexity.

\begin{lemma} \label{lemma:average_complexity}
The average sample complexity from \cref{thm:bias_refinement} per transmission $\rep \in [\nrep]$ is upper bounded by
\begin{align*}
    &\Ed{\samplecommon[1], \cdots \samplecommon[\nsamplescommon] \sim \pdfc, \repidxrow \sim \pdfauxc}{\nsamplescommon + \nsamplesindiv} \leq \mathcal{O}(\exp(\kl{\pdftc}{\pdfc})) \\
    &\!+ \! \frac{\chisq{\pdfc}{\pdftc}\!\!+\!\!1}{(\nsamplescommon-1)/\nsamplescommon} \Ed{\rvqc \sim \pdftc\!\!}{\mathcal{O}(\exp(\kl{\pdftcondcvar[\rvqc]}{\pdficondcvar[\rvqc]}))}\!\!.
\end{align*}
\end{lemma}
The communication cost is given by $\mathcal{O}(\kl{\pdftc}{\pdfc}) + \frac{\efrac}{(\nsamplescommon-1)/\nsamplescommon} \Ed{\rvqc \sim \pdftc}{\mathcal{O}(\kl{\pdftcondcvar[\rvqc]}{\pdficondcvar[\rvqc]})}$ bits.

\section{Multi-Terminal Setting} \label{sec:multi_terminal}

We next consider the case of multiple decoders, and for clarity, assume that $\nrec = 2$. The two decoders $\rec \in [2]$ observe samples from priors $\pdfi[1]$ and $\pdfi[2]$, that are marginal of a correlated joint distribution, thus $\mutinf{\rvi[1]}{\rvi[2]} > 0$. Recall that this joint distribution $\pdfjointtwo$ is known to all parties, but only the encoder can sample from both $\pdfi[1]$ and $\pdfi[2]$.

We first describe a straightforward, yet suboptimal, solution to the problem. The encoder draws a sufficient number of samples from the prior $\sampleindivonestd, \sampleindivtwostd \sim\pdfjointtwo$. Each decoder should be indicated $\nrep$ appropriate samples $\sampleindivstd[\repidx], \rep \in \nrep,$ through indices $\repidx$ to obtain an estimate $\isesti = \frac{1}{\nrep} \sum_{\rep = 1}^\nrep \function(\sampleindivstd[\repidx])$. To transmit the $\rep$-th sample, a naive strategy selects for decoder $1$ an appropriate sample index $\repidx[1]$ from $\sampleindivonestd[1], \cdots, \sampleindivonestd[\protect{\nsamplesrec[1]}]$, where $\nsamplesrec[1] = \mathcal{O}(\exp(\kl{\pdft}{\pdfi[1]}))$, and to decoder $2$ a sample index $\repidx[2] \in [\nsamplesrec[2]]$ to select a sample from $\sampleindivtwostd[1], \cdots, \sampleindivtwostd[\protect{\nsamplesrec[2]}]$, where $\nsamplesrec[2] = \mathcal{O}(\exp(\kl{\pdft}{\pdfi[2]}))$.

 Following the methodology of hierarchical sampling in the point-to-point case from \cref{sec:nested_sampling}, we instead propose to first select an appropriate sample from $\pdfc$ common to both decoders (transmitted through broadcast), and then select a sample from $\pdficondc[1]$ and $\pdficondc[2]$ individually for each decoder, transmitted by unicast.  This utilizes the availability of a broadcast channel and the non-zero Gács-Körner Theorem to reduce the communication cost. 
According to Gács-Körner \cite{gacs1973common}, if $\rvi[1]$ and $\rvi[2]$ have non-zero common information, then there exist two deterministic functions $\gkfunctions[1](\cdot)$ and $\gkfunctions[2](\cdot)$, defining a common random variable $\rvc = \gkfunctions[1](\rvi[1]) = \gkfunctions[2](\rvi[2])$ over the alphabet $\blockalphabet$. The Gács-Körner common information is then given by the $\rvc$ that maximizes the common information $$\cgk{\rvi[1]}{\rvi[2]} \define \max_{\rvc} \entropy{\rvc} \quad \text{ s.t. } \rvc = \gkfunctions[1](\rvi[1]) = \gkfunctions[2](\rvi[2]).$$ 
Given $\rvc$ with maximum entropy, the random variables $\rvi[1]$ and $\rvi[2]$ are independent. It is well-known that non-zero common information only exists if the joint distribution $\pdfjointtwo$ exhibits a block structure. The alphabet $\blockalphabet$ of $\rvc$ is then the blocks of the joint distribution. %
Having $\pdfjointtwo$, the encoder and both decoders can compute $\gkfunctions[i](\cdot)$, and thereby the probability law $\pdfc$ of the random variable $\rvc$. The encoder now samples from $\pdfjointtwo$, and the decoders observe samples from the marginals through shared randomness. Decoder $1$, by observing samples from $\pdfi[1]$, can thus obtain samples from $\pdfc$ by invoking $\gkfunctions[1](\cdot)$. Decoder $2$ draws the same samples for $\rvc$ by observing samples from $\pdfi[2]$ induced by the joint $\pdfjointtwo$ and invoking $\gkfunctions[2](\cdot)$. The samples from $\pdfc$ can be used to optimize the importance sampling strategy under dependent yet different priors.

We use as $\rvc$ in \cref{sec:nested_sampling} the common random variable resulting from the Gács-Körner common information over blocks of the joint distribution $\pdfjointtwo$ with alphabet $\blockalphabet$, such that the partitioning functions $\gkfunctions[1](\rvi[1])$ and $\gkfunctions[2](\rvi[2])$ map the priors $\rvi[1]$ and $\rvi[2]$ to $\rvc$. As in \cref{sec:nested_sampling}, we define a probability measure $\pdftc$ analog to $\pdfc$ over the alphabet $\blockalphabet$. For a low bias of the sampling (cf. \cref{thm:bias_refinement}), the number of common samples $\nsamplescommon$ from $\pdfc$ needs to be in the order of $\exp(\kl{\pdftc}{\pdfc})$. Following the methodology in \cref{paragraph:block_sampling}, the encoder draws $\nsamplescommon$ samples from $\pdfc$ to obtain $\samplecommon[1], \cdots \samplecommon[\nsamplescommon] \sim \pdfc$, by sampling from $\pdfjointtwo$ and invoking function $\gkfunctions[1](\cdot)$ or $\gkfunctions[2](\cdot)$ on the samples of $\rvi[1]$ or $\rvi[2]$. The encoder selects an index $\repidxrow$ by sampling from the weight distribution $\pdfauxc$ defined in \eqref{eq:pdfauxc}.
Conditioned on the common information $\rvc$, the priors are independent. The encoder draws samples from $\pdfjointtwocondc = \pdficondc[1] \pdficondc[2]$; hence for each decoder $\rec$ individually and each $\repidxrow \in [\nsamplescommon], \rep \in [\nrep]$, the encoder follows the methodology in \cref{paragraph:conditional_sampling} and draws $\nsamplesindivvar[\protect{\samplecommon[\repidxrow]}]$ samples $\sampleindiv[1], \cdots, \sampleindiv[\protect{\nsamplesindivvar[\protect{\samplecommon[\repidxrow]}]}]$ from the conditional $\pdficondc(\cdot \vert \samplecommon[\repidxrow])$. The encoder selects for each decoder an index $\repidxcol$ by sampling from $\pdfauxicvar[\protect{\samplecommon[\repidxrow]}](\sampleidx)$ as defined in \eqref{eq:pdfauxic}. 
Hence, the importance sampling estimate reads $\isesti = \frac{1}{\nrep} \sum_{\rep = 1}^\nrep \function(\sampleindiv[\repidxcol]).$

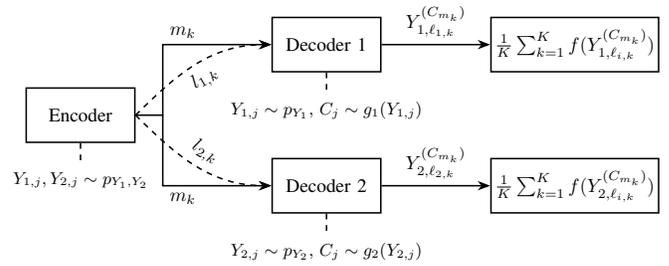
\begin{figure}[!t]
    \centering
    \resizebox{\linewidth}{!}{
    \begin{tikzpicture}[node distance=1.5cm and 3.5cm, >=Stealth, thick]

\node[draw, rectangle, minimum width=2cm, minimum height=1cm, align=center, inner sep=2pt] (encoder) {Encoder};

\node[draw, rectangle, minimum width=2cm, minimum height=1cm, align=center, right=2.5cm of encoder, yshift=1.3cm, inner sep=2pt] (receiver1) {Decoder 1};
\node[draw, rectangle, minimum width=2cm, minimum height=1cm, align=center, right=2.5cm of encoder, yshift=-1.3cm, inner sep=2pt] (receiver2) {Decoder 2};

\draw[->, thick] (encoder.east) -- ++(.5, 0) coordinate (midpoint) |- node[midway, above right, text=black] {$m_k$} (receiver1.west);
\draw[->, thick] (midpoint) |- node[midway, below right, text=black] {$m_k$} (receiver2.west);

\draw[->, dashed, thick] (encoder.east) to[out=45, in=180] node[midway, below, sloped, text=black] {$l_{1,k}$} (receiver1.west);
\draw[->, dashed, thick] (encoder.east) to[out=-45, in=180] node[midway, above, sloped, text=black] {$l_{2,k}$} (receiver2.west);

\node[below=0.4cm of encoder] (sample_encoder) {\small $Y_{1,j}, Y_{2,j} \sim p_{Y_1, Y_2}$};
\draw[-, dashed] (encoder.south) -- (sample_encoder.north);

\node[below=0.4cm of receiver1] (sample_receiver1) {\small $Y_{1,j} \sim p_{Y_1}, \, C_j \sim g_1(Y_{1,j})$};
\draw[-, dashed] (receiver1.south) -- (sample_receiver1.north);

\node[below=0.4cm of receiver2] (sample_receiver2) {\small $Y_{2,j} \sim p_{Y_2}, \, C_j \sim g_2(Y_{2,j})$};
\draw[-, dashed] (receiver2.south) -- (sample_receiver2.north);

\node[draw, rectangle, minimum width=3cm, minimum height=1cm, align=center, right=.5cm of receiver1, xshift=1.5cm, inner sep=2pt] (function1) {$\frac{1}{\nrep} \sum_{\rep=1}^\nrep \function(\sampleindivone[\repidxcol])$};
\node[draw, rectangle, minimum width=3cm, minimum height=1cm, align=center, right=.5cm of receiver2, xshift=1.5cm, inner sep=2pt] (function2) {$\frac{1}{\nrep} \sum_{\rep=1}^\nrep \function(\sampleindivtwo[\repidxcol])$};

\draw[->, thick] (receiver1.east) -- (function1.west) node[midway, above, text=black] {$\sampleindivone[\protect{\repidxcol[1]}]$};
\draw[->, thick] (receiver2.east) -- (function2.west) node[midway, above, text=black] {$\sampleindivtwo[\protect{\repidxcol[2]}]$};

\end{tikzpicture}}
    \vspace{-.5cm}
    \caption{Simple illustration of hierarchical sampling for two decoders with correlated priors $\pdfi[1]$ and $\pdfi[2]$ using Gács-Körner common information $\rvc$. \vspace{-.5cm}} %
    \label{fig:block_diagram}
\end{figure}

The common index $\repidxrow$ can be broadcast to both clients, and $\repidxcol[1]$ and $\repidxcol[2]$ are transmitted via point-to-point links. While the first sampling operation is common to all decoders, the latter (conditional) sampling can be regarded as the refinement for the individual priors. We depict in \cref{fig:block_diagram} a minimal schematic of our method.

When $\nsamplescommon \geq \exp(\kl{\pdftc}{\pdfc} + \slack_c)$ and $\forall \commonvar, \rec \in \blockalphabet: \nsamplesindiv \geq \exp(\kl{\pdftcondcvar[\commonvar]}{\pdficondcvar[\commonvar]} + \slack)$, the guarantee from \cref{thm:bias_refinement} holds uniformly for both decoders with probability $1-2(\vert \blockalphabet \vert (\maxblockeps[1]+\maxblockeps[2]) + 2\epsilon)$. According to \cref{lemma:average_complexity}, the average communication cost is given by $\mathcal{O}(\kl{\pdftc}{\pdfc}) + \frac{\efrac}{(\nsamplescommon-1)/\nsamplescommon} \Ed{\rvqc \sim \pdftc}{\sum_{\rec=1}^\nrec \mathcal{O}(\kl{\pdftcondcvar[\rvqc]}{\pdficondcvar[\rvqc]})}$.
The major advantage of this scheme compared to the standard scheme is that the cost introduced by the divergence of the distributions $\pdftc$ and $\pdfc$ over the blocks $\blockalphabet$ is only incurred \textit{once} per transmission, compared to incurring cost for each of the decoders. This can significantly reduce the communication cost for non-zero common information, and is amplified in the case of multiple receivers.

The choice of $\rvc$ as the entropy-maximizing common random variable does not directly relate to the KL-divergence between target distributions $\pdftc, \pdftcondc$ and the prior distributions $\pdfc, \pdficondc$. Even without any information about $\pdftc$ and $\pdftcondc$, $\rvc$ could be chosen to minimize the communication cost on expectation. Furthermore, the encoder can send refinements of the functions $\gkfunctions[1](\cdot)$ and $\gkfunctions[2](\cdot)$ to both decoders, in order to optimize the KL-divergence $\kl{\pdftc}{\pdfc}$ and $\kl{\pdftcondc}{\pdficondc}$ at the expense of a one-time overhead. This overhead vanishes with increasing $\nrep$. 
If the joint distribution $\pdfjointtwo$ was unknown to the decoders, it is similarly possible that the encoder determines desirable functions $\gkfunctions(\cdot)$ to minimize subsequent communication costs and transmits the partitioning of the alphabet given by $\gkfunctions(\cdot)$ to the individual decoders. The remainder of the method remains unchanged.

\begin{remark}
The concept of common information can be generalized to more than two random variables \cite{tyagi2011when}, facilitating a generalization of our methodology to the case of more than two decoders. Going beyond, the introduced hierarchical sampling strategy generalizes to more hierarchical levels, which can be leveraged in the case of many decoders. E.g., it is possible to first encode a common sample for all decoders, conditioned on which one can create common samples for subsets of the clients. This can be repeated until ultimately resorting to pair-wise common information. However, we note that such common information structures can be rare in practice.
\end{remark}

\section{Conclusion}
We introduced a hierarchical remote generation and estimation strategy based on MRC that leverages prior information shared between encoder and decoder to evaluate a given function over a target distribution through a rate-constrained communication link. We proved an error bound and compared it to the MRC baseline. We showed that hierarchical sampling can be beneficial in the point-to-point setting (depending on the function properties), and significant gains can be achieved for priors with non-zero common information. The scheme exhibits the structure of successive refinement, where the broadcast link is used to transmit a coarse sample, later refined for each decoder individually over unicast links. Extending our methodology to multiple hierarchies can facilitate the usage in successive refinement. The details of such a variant remain an interesting open problem, \camready{as well as studying alternatives to the Gács-Körner common information}.

\balance
\bibliographystyle{IEEEtran}
\bibliography{refs}

\end{document}